# Automatic Optimization for MapReduce Programs

Eaman Jahani
University of Michigan
Ann Arbor, MI 48109-2121
ejahani@umich.edu

Michael J. Cafarella
University of Michigan
Ann Arbor, MI 48109-2121
michjc@umich.edu

Christopher Ré
University of Wisconsin
Madison, WI 53709-1685
chrisre@cs.wisc.edu

## ABSTRACT
The MapReduce distributed programming framework has become popular, despite evidence that current implementations are inefficient, requiring far more hardware than a traditional relational databases to complete similar tasks. MapReduce jobs are amenable to many traditional database query optimizations (B+Trees for selections, column-store-style techniques for projections, *etc*), but existing systems do not apply them, substantially because free-form user code obscures the true data operation being performed. For example, a selection in SQL is easily detected, but a selection in a MapReduce program is embedded in Java code along with lots of other program logic. We could ask the programmer to provide explicit hints about the program's data semantics, but one of MapReduce's attractions is precisely that it does not ask the user for such information.

This paper covers MANIMAL, which automatically analyzes MapReduce programs and applies appropriate data-aware optimizations, thereby requiring no additional help at all from the programmer. We show that MANIMAL successfully detects optimization opportunities across a range of data operations, and that it yields speedups of up to *1,121%* on previously-written MapReduce programs.

## 1. INTRODUCTION

The MapReduce programming framework has become an extremely popular method for developing distributed data-processing software. It offers developers an environment that is quite different from that of a traditional relational database system, and more similar to standard UNIX development: software is written using traditional languages rather than SQL, data is treated as bytestreams rather than tuples, and there is no explicitly-declared metadata. Further, MapReduce systems have shown they can operate at extremely large scale with manageable administrative overhead: Yahoo has announced one that uses 10,000 cores [30]. Although the original motivation of MapReduce's designers was scalability for Web-scale bulk-processing tasks [12], some common MapReduce programs – such as simple selection and aggregation of log file data – overlap with traditional relational workloads.

However, MapReduce systems lag far behind RDBMSes in their query-processing sophistication and runtime efficiency. For example, Pavlo *et al.* [23] showed that a MapReduce program can run 2-50x slower than a similar relational query run on an RDBMS, using identical hardware. A recent paper by Anderson and Tucek [5] suggested that Hadoop performed bulk data processing at a rate of less than 5 megabytes per second per node (and barely more than half a megabyte per second per core) [5].

Thus, MapReduce systems may obtain a required level of performance through sheer scalability, but only at an enormous cost in hardware and power. If MapReduce could employ the efficient query processing techniques commonly found in RDBMSes, existing clusters could accomplish substantially more work, or could accomplish the same work with much less hardware. Developers would not have to choose between the appealing MapReduce programming model and the efficiency of an RDBMS; they could enjoy both.

Our MANIMAL system enables MapReduce program execution with substantial speedups over conventional systems. MANIMAL does so using identical hardware, and does not ask the developer to make *any* program modifications. We observed speedups of up to *1,121%* on *wholly-unmodified* programs published by Pavlo, *et al.* [23]. Our experiments evaluate MANIMAL using all of the benchmarks published by Pavlo, *et al.*, as well as programs we wrote to examine each individual optimization method.

Previous work in query optimization has mainly focused on operators in the relational algebra. Of course, the MapReduce framework (like most software in general) does not use this algebra. However, *data-centric programming idioms* often perform work that is similar to that done by the relational algebraic operators. For example, a MapReduce `map()` function that only emits data when a deserialized parameter's `rank` is greater than 1 performs work that is akin to a *selection*. A `map()` that only examines a subset of the fields in its passed-in parameters performs work that is similar to a *projection* of fields in the object. We can apply known query optimization techniques to speed up such code. For example, in the case of a projection-style MapReduce program, the system can use an alternate serialized version of the data that stores only the needed fields for a program, thereby reducing the overall number of bytes that must be processed (similar to a column-store [26] or an on-disk binary association table [8]). The goal of MANIMAL is to automatically detect and exploit as many of these standard optimization opportunities as possible.

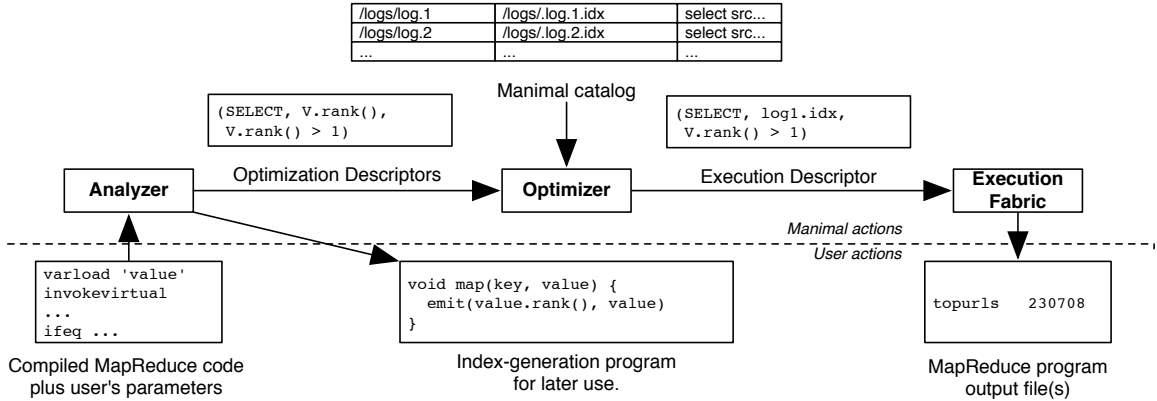

Figure 1: Architecture of the MANIMAL system.

Of course, relational-style optimizations are tightly linked to the semantics of the program itself. RDBMSes use query languages and metadata that make such semantics explicit, but MapReduce systems do not. Moreover, programmers likely chose to write a MapReduce program at least partially *because* such metadata is not required. Thus, a central challenge for MANIMAL is to understand users' programs well enough so that the optimizations can be applied wholly automatically, thereby obtaining good performance and preserving MapReduce's appealing programming model.

This paper describes a static analysis-style mechanism for detecting optimizable code in already-compiled MapReduce programs. Like most programming-language optimizers, it is a *best-effort* system: MANIMAL does not guarantee that it will find every possible optimization, and a determined programmer can elude the detector. Of course, missing an optimization is regrettable, but finding a false one is catastrophic; MANIMAL should only indicate an optimization when it is entirely safe to do so. The MANIMAL analyzer is designed to sacrifice potential optimizations when there is a chance of non-safety; however, we show experimentally that it can find most of the optimizations discovered by a human annotator in a collection of MapReduce programs.

Note that this static analysis approach is most appropriate for MapReduce programs that are "program-specific," with code that is directly related to the user's end-goals for the program. For tools layered on top of MapReduce, such as the Pig query system [22], we believe a better approach is for the tools to give MANIMAL explicit hints about program semantics. We discuss this issue in more depth in Appendix A.

**Background** There has been a recent surge of interest in MapReduce systems. Some projects have applied MapReduce-inspired techniques to building a traditional relational database [2, 27], but most have focused on improving MapReduce execution performance [3, 9, 13, 28, 29]. However, most of these projects are either low-level system techniques that are semantics-free, or ask the user to modify their code to expose more program semantics. To the best of our knowledge, MANIMAL is the first MapReduce system to use data semantics-driven optimizations without requiring *any* code changes from developers.

We previously presented an outline of the MANIMAL architecture and a single experimental result [10]. This paper substantially expands on that earlier work, with new optimization techniques, a much more detailed technical discussion, more complete discussions of MapReduce workloads, and full experimental results.

**Contributions and Outline** The main contributions of this work include:

- A framework for optimizing wholly-unmodified MapReduce programs, targeting data-centric programming idioms (in Section 2)

- Algorithms for detecting and exploiting three optimization types: selection, projection, and data compression (Section 3).

- An implemented MANIMAL system that yields substantial performance gains (up to more than 11x) on the previously-published benchmark programs from Pavlo, *et al.* (Section 4).

Finally, we conclude with a discussion of related work in Section 5. We also discuss MapReduce-related tools in Appendix A and summarize the state of MapReduce benchmark workloads in Appendix B. We present additional information on program analysis and experimental results in Appendices C and D. We discuss ideas for future work in Appendix E.

## 2. SYSTEM OVERVIEW

MANIMAL comprises three main components that allow it to optimize a user's MapReduce program wholly automatically. Figure 1 shows the flow of information through the system. The **analyzer** examines a user's submitted MapReduce program and sends the resulting *optimization descriptor* to the **optimizer**. The **optimizer** uses this descriptor, plus a catalog of precomputed indexes, to choose an optimized execution plan, called an *execution descriptor*. This descriptor, plus a potentially-modified copy of the user's original program, is then sent to the **execution fabric** for full execution on the cluster. The **execution fabric** retains the standard *map-shuffle-reduce* sequence and is almost identical to standard MapReduce (in the case of our prototype, Apache Hadoop).

The vast majority of MANIMAL is hidden from the user. The submitted program does not need to be modified by the programmer in any way, and the final program output should be the same as what would have been generated by a conventional MapReduce system. The user should be able to detect just one difference (aside from improved runtime performance): submitting a job for execution yields not just a program result, but also an *index-generation program*. This program is itself a MapReduce program, and when executed generates an indexed version of the submitted job's input data. This indexed version of the data is used by the **optimizer** when selecting a best course of action. The decision to run an index-generation program is left to the system administrator, much like the decision to create an index in an RDBMS.

For the moment, we focus only on single MapReduce programs, not chains of such programs in which the output of one phase forms the input of another. However, we would like to address such pipelines in the future, as we discuss in Appendix E. Further, MANIMAL currently focuses mainly on relational-style programs. We have not yet examined some important classes of MapReduce tasks, such as inverted index construction. This decision partially reflects the availability of existing MapReduce workloads (we discuss workloads in detail in Section 4.1 and Appendix B). However, we believe the general **analyzer**-driven MANIMAL framework could in the future be applied to other program types, such as text-processing and iterative numeric hill-climbers.

We will now provide some background on each of the optimizations MANIMAL currently pursues. We will then give a brief system walkthrough from the user's perspective.

## 2.1 Optimizations

MANIMAL looks for three different kinds of optimizations.

**Selections** appear in MapReduce code if map() emits data only when a parameter-dependent conditional test holds true. For example, the following map() function only emits data when the input WebPage value has a rank of more than 1:

```
void map(String k, WebPage v) {
  if (v.rank > 1)
    emit(k, 1);
}
```

This code has the effect of of filtering elements from the input data. As with a relational selection, there is no point in executing this code for inputs that will fail the test; any invocation of map() for data that fails the conditional test is in a sense wasted work. And just as with a relational database, we can optimize such code at runtime by using a B+Tree to scan just the relevant portion of the input data. (MANIMAL uses the index to skip map invocations that do not yield output data, even if doing so may also mean skipping generating messages for the debug log.)

Other work has applied selection-appropriate indexing techniques to MapReduce [13]. The critical contribution of MANIMAL is to detect these selections automatically in unmodified developer code. MANIMAL currently uses a B+Tree, but in the future could also employ an R-Tree [16] or some other indexing technique when appropriate.

**Projection** optimizations modify the on-disk data file to only store bytes that are actually necessary for executing the user's code. For example, the user's code may take a WebPage as an argument, but may never examine the large htmlContent field. Eliminating unnecessary fields from the file reduces the total number of bytes that must be processed by MapReduce without changing the program's behavior. MANIMAL can apply projection optimizations because it can examine user code to see which fields are actually critical, and can examine the serialized class to see which fields are present. Standard MapReduce cannot automatically determine which bytes are safe to remove.

This optimization is similar to a simplified version of a column-store [26]. In the future we could modify MANIMAL projection to use "column-groups" that break input data into different smaller files, increasing the number of user programs that could use an index, at the cost of possibly-increased program execution time.

**Data compression** is different from Hadoop's built-in compression support. Hadoop stores a compressed version of input data on disk and of intermediate map data. Hadoop decompresses the data immediately prior to map() and reduce(). Hadoop can use any of several compression techniques, but in all cases applies the technique to the entire data file. Instead, MANIMAL enables two semantics-aware forms of compression, both previously used by Abadi, *et al.* [1]:

First, **delta-compression** efficiently stores runs of numeric values, by only keeping differences between values, instead of the absolute values. Storing just small deltas, when combined with a size-sensitive representation, can yield large storage savings. Standard MapReduce cannot apply this technique: the system must know which bytes are in the same field and are numeric. MANIMAL can discover this information via map()'s serialized input classes.

Second, **direct-operation** potentially allows the system to operate directly on compressed values. For example, a url that is used only in equality tests does not really need to be decompressed prior to map(); it is possible to use a compressed version of the url that preserves equality testing, and thus still yields the correct program output.[1] The program saves time because the compressed data is smaller than it would otherwise be, the data does not need to be decompressed prior to processing, and operations on the data may be faster. Direct-operation is possible because MANIMAL can determine when a potentially-compressed value would be used strictly in safe settings.

## 2.2 Walkthrough

A MANIMAL user submits a job for processing just as with conventional MapReduce. She provides:

- The compiled program and libraries, plus metadata like the name of the class that contains main().

- The input filename(s), plus program parameters and relevant configuration information

**Step 1 – Analysis**. The **analyzer** examines the input, in particular the compiled program, and attempts to detect optimization opportunities.

---

[1] MANIMAL can use direct-operation compression on the map()'s output key as long as the user does not require the final program output to be in sorted order.

The **analyzer** uses static program analysis techniques. These techniques make up MANIMAL's most novel intellectual contributions, and are described in detail in Section 3 below. The resulting *optimization descriptor list* has, for each applicable optimization, a label that identifies the optimization and optimization-specific parameters. For example, the `SELECT` descriptor includes includes a description of which values should be indexed, plus a logical formula over these values that describes when the `map()` may emit data. In Figure 1, the indexed value is `V.rank()`, and the function only needs to run when `V.rank() > 1`.

This component also creates an *index generation program* that runs on the same input data as the user's program. For example, the selection optimization yields a program that creates a B+Tree-based index. When indexes are eventually combined with the optimization descriptors, MANIMAL has all the information needed to invoke `map()` only when needed, thereby avoiding substantial amounts of work. Each run of an index generation program is tracked in the filesystem **catalog**.

Of course, MapReduce inputs are flat files, not structured tables, so any notion of indexing is somewhat surprising. But such files are usually a series of serialized objects, the names of which can be obtained via the declared types for the user's `map()` function and other supporting code. The program above operates on files that consist of serialized `String` and `WebPage` pairs. The code that serializes and deserializes these classes effectively declares the file's schema.

Put another way, output from the **analyzer** is akin to a description of a view on the data from the user's input file, which is materialized by the index generation program, and processed by the user's MapReduce program.

Anything that does not impact the program's final output is fair game for the **analyzer** to consider for downstream removal or modification, including code that has side effects such as debugging statements, network connections, and filewrites. MANIMAL can currently detect, though not optimize, such side effects.[2]

Indexes that arise from the **analyzer**'s work carry cost in disk space and in computational time to be created. Thus, as with relational data, MANIMAL indexing is not worth the effort for ephemeral "read-once" data files that are thrown away after processing just once. We believe this is acceptable for many MapReduce-related datasets, especially when one considers that different parties may analyze the same raw data: web access logs, crawls, *etc.*

The **analyzer** may find several orthogonal optimization opportunities in a given program, and so could emit several possible different index-generation programs. For example, a program that would benefit from both selection and projection could make use of several different indexes: one version that supports selection, one that supports projection, or one that supports both. The "best" index to compute depends partially on the system's index space budget and partially on the expected future workload. For example, it may not be wise to compute the combined selection-projection index if 9 out of 10 future jobs do no projections. MANIMAL does not yet attempt to address this issue in a disciplined way; the current **analyzer** always chooses the index program that exploits as many optimizations as possible.

---

[2]It would be possible to add a MANIMAL "safe mode" that avoids optimizations that modify side effects, at the possible cost of reduced optimization opportunities.

**Step 2 – Optimization**. The **analyzer** sends its list of optimization descriptors to the **optimizer** component. The **optimizer** examines the descriptors, the user's input file, and the **catalog** to choose the most efficient execution plan currently possible. The resulting *execution descriptor* indicates to the final **execution fabric** which index file to use, and which optimizations should be applied.

The **optimizer** faces two planning questions which in the long run should be determined by a cost-based approach, but for now are solved with simple rule-based heuristics. First, some MANIMAL optimizations may not be combinable for low-level implementation reasons; the optimizer may need to decide between multiple options.[3] Second, the **optimizer** may have to decide among several indexes compatible with the submitted program. It currently decides using a simple hard-coded ranking of applicable optimizations.

**Step 3 – Execution**. Finally, the **execution fabric** uses the execution descriptor to actually run the program. Most of the **execution fabric** is identical to a traditional MapReduce system. Our Hadoop-based prototype has a few modifications to support B+Tree-indexed input formats and deltacompression. The other optimizations – column-oriented files to support project, and direct operation on compressed data – can be performed without any infrastructure-level support at all.

## 3. ANALYZER IN DEPTH

The architecture described in the Section above is clearly inspired by the traditional RDBMS query-optimization-and-execution loop. However, MapReduce systems to date have rarely attempted any semantics-aware optimization at all; user programs have been treated as wholly-opaque binary objects. The **analyzer** is what moves user code into the realm of identifiable data operations, and thus is absolutely critical to MANIMAL's ability to provide *automatic* MapReduce optimization.

We view this ability to apply *existing* optimizations as MANIMAL's core contribution. While many published MapReduce research projects have found possible performance improvements, we are unaware of any that have been broadly deployed; we believe MANIMAL can be a "force multiplier" that eases many MapReduce optimization techniques into wider use.

MapReduce languages are usually Turing-complete ones, such as C and Java, and it is not *a priori* obvious that a successful **analyzer** can be built. For example, consider the daunting level of analysis required to apply HaLoop [9], which optimizes gradient-ascent-style MapReduce programs that iterate repeatedly over a dataset. An **analyzer** that detects such code would have to track data flows across multiple individually-launched MapReduce jobs. A user's MapReduce code for join, which could be improved by using a proper join algorithm, can also be very complicated: it may consume `map()` data from multiple files, in which each joined table's data comes from a different source file. Our current **analyzer** successfully addresses a subset of possible optimizations: the above-listed selection, projection, and compression approaches.

The static analysis approaches employed by the **analyzer**

---

[3]Our current optimization set presents just one conflict; we currently favor selection over delta-compression.

are well-known in the programming languages and compiler community [4]. In this section we provide a brief background on static analysis techniques for *control flow analysis* and *data flow analysis*. We then describe how we use these tools to construct the MANIMAL **analyzer**.

### 3.1 Background: Static Analysis

Static analysis uses code inspection, rather than any kind of live program instrumentation, to learn about a piece of software. The analysis techniques used by MANIMAL are well-known and can be found in compiler texts such as Aho, *et al.* [4]. One of the most basic tools for control flow analysis is the *control flow graph* (CFG). A CFG for a method contains a node for each block of statements, and directed edges that represent control transitions from one block to another. A sequence of statements that employ no control-flow primitives yield a region of code with one entrance and one exit point; such sequences can be merged into a single *basic block*. Therefore, the CFG encapsulates all possible paths the program might take during execution. Edges lead from each node in the graph to all potential immediate successors. We also create two special nodes, one for "function entry" and one for "function exit." Figure 4 shows the CFG for the `map()` function from Section 2.

Another technique we use is *dataflow analysis*, in particular the computation of *reaching definitions*. The *definition* of a variable (that is, an assignment to the variable) at statement $d$ is said to *reach* a *use* of that variable at statement $u$, as long as $u$ is reachable from $d$ in the CFG, and there is no intervening definition for the variable between $d$ and $u$. A *use-def* chain for a variable consists of a use of the variable at statement $u$, and all reaching definitions of the variable. We thus construct a node for $u$, with an edge pointing from $u$ to each definition node $d$ that can be reached without any intervening definition. Figure 5 shows an example of the *use-def chains* computed for various statements in the `map()` in the previous Section.

### 3.2 Mechanism: Selection

Our current **analyzer** searches for several traditional optimizations: selection, projection, and compression. It does so at the "micro-scale," and only optimizes the `map()` function. (We plan to examine `reduce()` in future work.) It thus resembles single-function optimization more closely than inter-method analysis. Such an approach works well because for these MapReduce code idioms in observed code are small and mainly fit in a single function. In this section we describe how analysis works for selection. We cover projection and data compression in Appendix C.

Much of the **analyzer** machinery focuses on ensuring that all discovered optimizations are *safe* – that is, optimizations that observe the semantics of the original program and produce the same outputs from the reduce stage.

The MANIMAL **analyzer** takes as input the compiled Java class files that contain the target MapReduce job, plus basic user parameters such as the input file, *etc.* We use the ASM bytecode manipulation library to process the compiled program [7]. Although the **analyzer** is currently implemented for Java, the same principles apply for any compiled language in which the CFG and *use-def* chains can be accurately computed.[4]

---
[4]Languages that permit arbitrary pointer arithmetic and

```
void map(String k, WebPage v) {
  numMapsRun++;
  if (v.rank > 1 || numMapsRun > 200)
    emit(k, 1);
}
```

**Figure 2: The flow of control for this function relies on member variable `numMapsRun`. Because output may not strictly depend on the function's parameter inputs, MANIMAL cannot safely optimize this code.**

The **analyzer** starts by examining `map()` for optimization opportunities. The algorithm for selection appears in Figure 3. There are several auxiliary functions. $isEmit(s)$ tests whether a statement $s$ *emits* data to the `reduce()` step. $paths(s)$ returns all possible CFG paths that reach $s$. $conds(s)$ returns all conditional statements that appear in the given CFG path.

In addition, we define two methods – $getUseDef(s)$ and $isFunc(useDefChain)$ – to test whether optimizations are safe. The former computes a *use-def DAG* for $s$. $getUseDef()$ starts as a single *use-def* chain, but for each *def* node, **analyzer** treats the *def* as a new *use* and recursively obtains its *use-def* chain, bottoming out when the *uses* have no more dependent *def* statements inside the `map()`. This can happen when a use depends on passed-in parameters or externally-defined member variables. The result is a directed acyclic graph that represents all the points in the `map()` that might influence the value of the initial statement $s$.

The functional test $isFunc(useDefChain)$ succeeds when all of the following hold:

- The *use* depends only on `map()` parameters or constants, not class members or other external variables.

- The *use-def DAG* contains no calls to methods which themselves may not be functional in terms of their inputs. That is: we must be careful the `map()` does not simply push its dependence on a class member off to some other method. The **analyzer** has built-in knowledge of standard language operations and some common class library methods, such as those associated with `String`, `Pattern`, *etc.*

A functional chain from input parameters to tuple-emission means that `map()`'s output is entirely determined by the input record, guaranteeing MANIMAL that its optimization decisions will be safe as long as the map inputs are preserved. As an example of code that fails the $isFunc(useDefChain)$ test, consider the code in Figure 2. It shows a `map()` whose output depends on both `v.rank` and the member variable `numMapsRun`. MANIMAL cannot optimize this code, because the indexed values that MANIMAL reads from disk and passes to `map()` are no longer sufficient for determining the method's output. If MANIMAL were to optimize the program by using a B+Tree to avoid unproductive invocations of `map()`, it would have the unintended side-effect of changing the value of `numMapsRun`, and thus also change data emit decisions. The **analyzer** considers programs like the one in Figure 2 to be unsafe and unoptimizable.

We can now examine the selection analysis algorithm in detail. The primary goal is to compute a logical formula over `map()`'s variables and input parameters that evaluates to

---
jumps can induce control flow patterns that cannot necessarily be detected using static analysis.

```
1: function findSelect(mapperStmts):
2:   allFunc ← true, condPaths ← {}, dnf ← false
3:   for s ∈ mapperStmts do
4:     if isEmit(s) then
5:       for all path ∈ paths(s) do
6:         condPaths ← condPaths ∪ conds(path)
7:         dnf ← dnf OR conj(conds(path))
8:   for all condPath ∈ condPaths do
9:     for all cond ∈ condPath do
10:      if not isFunc(getUseDef(cond)) then
11:        allFunc ← false
12: if allFunc return dnf else return {}
```

**Figure 3: The analyzer detection algorithm for selection.** $conj()$ **returns a conjunction of the logical conditional expressions in its input.**

true if and only if the function emits a tuple. In particular, the selection algorithm constructs a conditional statement in disjunctive normal form, in which there is a disjunct for each unique path to an *emit()* statement. Each of the disjuncts contain a conjunction of the conditional tests that must hold true to reach the `emit()` through its respective path. The algorithm is presented in Figure 3.

## 4. EXPERIMENTS

We ran three types of experiments to show that MANIMAL can effectively optimize users' programs. First, we show that the **analyzer** can detect most of the optimization opportunities in users' code. Second, we show that MANIMAL obtains substantial runtime improvements on these programs. Third, we tested the performance gain to be found using each individual optimization type; we discuss selection results briefly here, and discuss full results in Appendix D.

All of the experiments in this Section were performed using a small 5-node cluster, running Hadoop version 0.20.1. We used benchmark programs and data published as part of the work by Pavlo, *et al.* [23]. Result times are averaged over 3 runs.

### 4.1 Analyzer Recall

For MANIMAL to be broadly useful, its **analyzer** must be able to detect optimization opportunities in real MapReduce code. Here we test the **analyzer**'s recall on four benchmark programs made available as part of the work in Pavlo, *et al.*.

Unfortunately, while MapReduce has become very popular, the number of open-source MapReduce programs that we can examine is relatively small. Although the programs from Pavlo, *et al.* are well-known, simple, and downloadable, they are not ideal. In particular, they may overrepresent database-style operations and underrepresent text-centric and numeric processing. However, the tasks do fit reasonably well with the results of a recent survey that showed business data analysis and log processing to be the most popular Hadoop applications (ahead of Extract-Transform-Load and scientific applications) [18]. Indeed, these programs have already been used for evaluation purposes by Abouzeid, *et al.* [2] and Dittrich, *et al.* [13]. We discuss our choice of workload in more detail in Appendix B.

Table 1 shows the results of these experiments. We do not show results for direct-operation, as none of the tested benchmark programs permitted this optimization. For each cell in the table, we show whether the optimization was successfully *Detected*, or went *Undetected*, or was simply *Not Present*. A human observer examined the programs to see which optimizations were *present*. The **analyzer** emits no false positives. It fails to detect just three optimizations:

- **Benchmark 1, projection and delta-compression** fail because the authors employed an unusual custom class for the `map()` function's `value` parameter. The `AbstractTuple` class essentially creates its own serialization format, and contains no direct program-specific clues as to its function. The **analyzer** is thus unable to distinguish between different fields in the serialized data. This method of serializing data is inefficient and surprising; if the class were rewritten to employ standard topic-specific serialization, we could detect the optimization.

- **Benchmark 4, selection** fails because the code employs a Java `Hashtable` as part of the filtering process. The current version of MANIMAL does not have builtin knowledge of how `Hashtable` works, and so cannot tell that testing for a key in the Hashtable will only succeed if it had been inserted previously. However, `Hashtable` is a very commonly-used class, and adding custom handling of it would not be unreasonable.

Overall, we view the **analyzer** as successful. Not only does it pick up most available optimizations, two of the three mistakes are due to a programming approach that is likely to be fairly unusual. Moreover, the missed optimizations in the case of Benchmark 1 are likely to have little impact: delta-compression cannot be combined with selection, and experiments showed that gains from the missed projection opportunity were undetectable next to the huge *selection* gains.

The only serious optimization overlooked by MANIMAL is the selection condition in Benchmark 4. Note that one reason it is difficult to detect is that the code is the most text-centric of any of the Benchmarks; it does not directly map to a relational-style operation. Such programs are exactly where the performance gap between conventional MapReduce and RDBMSes is already relatively small, as Pavlo, *et al.* showed.

### 4.2 End-to-end Performance

We now evaluate MANIMAL's overall end-to-end runtime improvement on the above benchmarks. Of course, the runtimes here only reflect optimizations that were detected by the **analyzer**, described above. As much as possible, we chose our experimental parameters to match those in Pavlo, *et al.* [23]. We used the same code, used the same tools to generate test data, and used similar selectivities. However, the number and quality of our cluster machines differ, and so the data sizes here are unavoidably different. It is therefore inappropriate to directly compare our absolute execution times to those from the previous paper.

Experimental results are shown in Table 2. The MANIMAL programs described used only the detected optimizations as described in Section 4.1. All end-to-end times are averaged over three runs. In all of these cases, the time necessary to compute the index is at most 50% more than the time needed to execute the original Hadoop program; thus, even files that are examined infrequently can potentially benefit from MANIMAL optimization. All reported times include

| Test | Description | Select | Project | Delta-Compression |
|------|-------------|--------|---------|-------------------|
| Benchmark-1 | Selection | Detected | *Undetected* | *Undetected* |
| Benchmark-2 | Aggregation | Not Present | Detected | Detected |
| Benchmark-3 | Join | Detected | Not Present | Detected |
| Benchmark-4 | UDF Aggregation | *Undetected* | Not Present | Not Present |

Table 1: The results of running the MANIMAL analyzer on various MapReduce programs.

Hadoop startup and standard serialization costs. Any future efforts to reduce startup or serialization time (*e.g., Avro* [6]) are likely to result in better runtimes.

**Benchmark 1** By employing a selection index on Benchmark 1, MANIMAL can obtain a greater than 11x speed increase over standard Hadoop. We ran this benchmark with a threshold chosen to obtain the same selectivity as Pavlo, *et al.*: 0.02%. As mentioned above, the **analyzer** fails to detect a potential projection optimization for this task. However, the original file contains relatively few fields, and performing the projection yields an index that is smaller than the non-projected index by just 5.5%, and results in no further detectable performance improvements. It is interesting to note that MANIMAL's performance relative to Hadoop for this task is roughly midway between what Pavlo, *et al.* reported for their "DBMS-X" and Vertica results.

**Benchmark 2** Benchmark 2 offers opportunities for both projection and delta-compression optimizations, and MANIMAL detects both. Because so much of the original input file does not need to be read in order to produce the final result, the index is fairly small: 20% of the original file's size.[5] MANIMAL can run an optimized version of the task in roughly 1/3 of the time required by Hadoop. MANIMAL's running time relative to Hadoop is similar to the difference reported by Pavlo, *et al.*, when comparing Hadoop to both DBMS-X and Vertica.

**Benchmark 3**'s join operation yields very interesting results. Unlike standard relational databases, MANIMAL has absolutely no knowledge of join processing. However, the `map()` task for this benchmark imposes a selection predicate that removes all but 0.095% of the *UserVisits* data from consideration. By recognizing the selection, and only scanning the records that can pass this filter, MANIMAL can hugely reduce the number of bytes that pass through the overall processing pipeline.

The impact is substantial: MANIMAL obtains a speedup of 6.73x. This number is not as good as what can be obtained by systems that are join-aware, and MANIMAL's performance relative to Hadoop does not approach the 18x gains seen by Pavlo, *et al.*. In the future, it may be possible for the **analyzer** to detect joins, by testing whether conditional tests in `map()` effectively mirror the differences between tuples from multiple files sent to the MapReduce job. Doing so would enable MANIMAL to implement distributed join-processing and possibly obtain similar gains.

## 4.3 Individual Optimizations

We also show MANIMAL's effectiveness in obtaining improved MapReduce performance using each of the individual optimization types. Here we present just the selection results. Other results, and details on the dataset used, are presented in Appendix D.

For selection, we examine MapReduce programs that implement the following query:

```
SELECT pageRank, Count(url) FROM WebPages
WHERE pageRank > Threshold GROUPBY pageRank
```

where `Threshold` is chosen to yield one of various selectivities. (The full results are seen in Table 3 in Appendix D.) The `WebPage` objects are generated to roughly match real-world Web conditions, as described in Appendix D. Of course, for this experiment we examine *only* the selection optimization, even though others may apply to the query. These selectivities admit much more output data than those in our end-to-end experiments from Section 4.2. Performance is roughly linear with selectivity; for selectivities between 60% and 10%, MANIMAL obtains speedups between 1.59x and 7.10x.

**Summary** We have found that across four real-life MapReduce programs designed to model real tasks, MANIMAL automatically obtained substantial speedups for three of them. In two of these cases, it obtained performance gains roughly commensurate with those obtained by a traditional relational system described in Pavlo, *et al.*, and we obtained still-substantial speedup on a third. While MANIMAL's **analyzer** failed to discover potential optimizations, test runs showed that the benefits of applying them would have been negligible. Finally, we have shown that the **analyzer** can effectively discover data semantics in natural MapReduce programs, and can yield substantial performance gains.

## 5. RELATED WORK

There has been a large amount of recent work on MapReduce [3, 9, 13, 28, 29], though none that takes system's wholly-automated approach to optimization. Several efforts have explored the problem of scheduling task execution [17, 29]. Afrati and Ullman [3] investigated how to efficiently perform joins using MapReduce. Yang, *et al.* [28] extended the programming model to *Map-Reduce-Merge*, allowing the user to express different join types and algorithms.

HadoopDB [2] attempts to combine relational and MapReduce qualities into a single system. However, HadoopDB is designed to be a scalable parallel relational database; it uses Hadoop internally but does not optimize MapReduce programs.

There have been several recent index-style attempts to improve MapReduce performance, such as Dittrich, *et al.* [13]'s Hadoop++ system, and the column-oriented work of Floratou, *et al.* [15]. The former requires explicit support from the programmer, while the latter only requires physical storage reorganization; both could be used as targets for MANIMAL.

---
[5]We ran the "standard" version of the benchmark, which sums revenues for unique IP addresses, not the subnet-oriented version.

| Test | Description | Space Overhead | Hadoop | MANIMAL | Speedup |
|---|---|---|---|---|---|
| Benchmark-1 | Selection | 0.1% | 429.78 secs | 38.35 secs | **11.21** |
| Benchmark-2 | Aggregation | 20% | 5,496.29 secs | 1,855.65 secs | **2.96** |
| Benchmark-3 | Join | 11.7% | 6,077.97 secs | 903.752 secs | **6.73** |
| Benchmark-4 | UDF Aggregation | 0% | N/A | N/A | **0** |

Table 2: Overall performance improvement provided by MANIMAL, across the Pavlo benchmark tasks.

MANIMAL's analyzer employs compiler techniques for database-style optimizations. Chilimbi, *et al.* [11] tried to reorder *in-memory* data representations to improve cache behavior, a systems-level improvement suggested by program semantics. They did not examine disk-based approaches. MANIMAL has some qualities in common with work optimizing XQuery (*e.g.*, Ré, *et al.* [25]). In particular, MANIMAL follows the same general approach of using static analysis to automatically apply well-known optimizations to a novel language. There has also been work in integrating data-manipulation primitives with traditional programming languages, as in LINQ [20]. Such tools side-step the need for MANIMAL's analyzer but also require the programmer to use a potentially-novel programming language.

# 6. CONCLUSIONS

We have described the MANIMAL system for optimizing MapReduce programs. MANIMAL automatically obtains substantial speedups, ranging from 296% to 1,121% on real MapReduce code. Moreover, MANIMAL provides an appealing framework for deploying many other achievements in MapReduce-optimization research.

# 7. ACKNOWLEDGMENTS

This work was supported by National Science Foundation grant IIS-1054009 as well as gifts from Google, Johnson Controls, and Yahoo!.

# APPENDIX
## A. LAYERED TOOLS

Certain popular tools, such as Pig, Hive, and Mahout, are layered on top of the MapReduce infrastructure. Users of these tools generally never see the MapReduce programmer interface, but instead use a tool-specific language. The tools then implement the user features with a MapReduce job (or set of jobs) that is executed on a MapReduce cluster. For example, Hive processes queries written in SQL, using MapReduce as the distributed query execution infrastructure. It would be very possible for these systems to be implemented on top of a different execution system.

The actual MapReduce source code for Pig and Hive jobs essentially work as interpreters for user code, and do not reflect the semantics of the user's program. This level of indirection makes it extremely difficult, if not impossible, for a static analysis-based tool to obtain performance gains from program-specific semantics. Similarly, Mahout tasks consume a generic textual data format that is designed to model any domain: it reflects nothing about the user's task at hand and so cannot be easily optimized.

Some anecdotal evidence suggests these jobs comprise a large and growing percentage of total MapReduce work. A recommended list of Hadoop "best practices" from a prominent Yahoo engineer suggested that raw Java tasks only be used when "really necessary." [21] A recent survey of Hadoop users [18] found that over the coming year users on average plan to increase their use of MapReduce-layered tools (*e.g.*, Hive from 44% to 52%; Mahout from 14% to 24%) while decreasing their use of raw Java (from 86% to 74%). We confirmed this trend in personal discussions with executives at the Hadoop company Cloudera, who suggested that Hadoop users show a pattern of migrating to higher-level tools over time, and away from lower-level Java programs [24]. Of course, there is still a substantial set of MapReduce jobs that these tools do not handle, especially those that involve lots of user-written code or process raw text or parse-heavy log formats.

Luckily, such tools usually have access to a very high-level description of the user's desired job semantics, which can be even better information than what can be recovered from a traditional MapReduce program using the MANIMAL **analyzer**. For such tools, MANIMAL is designed to be able to sidestep the **analyzer** and accept optimization descriptions directly, allowing the resulting synthesized jobs to still make use of the MANIMAL physical optimizations.

## B. BENCHMARKS AND WORKLOADS

Evaluating MANIMAL's practical impact is difficult, substantially because there is no agreed-upon workload of MapReduce programs. In addition to the layered tools described above, some candidate workloads include the following.

- **MapReduce task collections** such as those suggested by Dean and Ghemawat, including distributed grep, wordcount, and counting URLs in an access log. These are uniquely tailored for MapReduce, as many of them would be difficult using an RDBMS. However, they also very text- and graph-heavy, and are skewed toward Web search tasks. MapReduce has obtained wide popularity beyond its Web search beginnings, and it would be surprising if most of these tasks (except for the access log) were broadly representative today. Web page and graph-processing applications were not listed as a motivation for Hadoop users in a recent survey [18] (except perhaps as part of the "Other" category, which appealed to 8% of respondents).

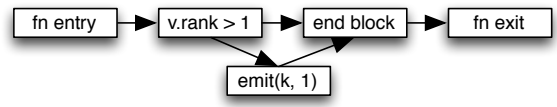

Figure 4: The control flow graph for the function in Section 2. Nodes represent basic blocks. Edges represent possible control flow paths.

- **The Gridmix workload** [14] was designed and used by Yahoo! as a low-level Hadoop performance testing tool. Is not designed to be a workload that is representative of a broader class of jobs. Instead, it is a pure "byte-level task" designed to stress and consume cluster resources in ways that correspond to a recorded cluster workload. For example, a Hadoop cluster that is running the Gridmix workload and a cluster that is running the recorded productive workload should exhibit the same number of `map()` and `reduce()` tasks per job, the same number of input and output bytes per job, the same access distribution over file blocks, the same job inter-arrival interval, *etc*. The actual work performed by Gridmix is meaningless: it simply consumes and emits random bytes according to recorded parameters. However well Gridmix may exercise the underlying Hadoop implementation, without any true task semantics to analyze, there is nothing MANIMAL can do to improve its execution.

- **TPC-H** is a standard benchmark for evaluating the performance of report-generation queries in RDBMSes, and has been used by Kim, *et al.* [19] to evaluate MapReduce. However, TPC-H has no special affinity with MapReduce and its queries do not exercise many of its interesting features, such as easy text parsing.

- **Analytical tasks** described by Pavlo, *et al.*. Beyond the original grep task, these include selection, aggregation, join, and UDF-driven aggregation. These are designed to roughly model a log- and crawl-processing workload. One possible weakness of this suite is that it consists of just a few programs, and does not attempt to emulate any well-documented workload.

Thus, our current approach is to study end-to-end MANIMAL performance on the tasks suggested Pavlo, *et al.*. These programs are diverse in terms of potential optimizations, appear to fit MapReduce practice reasonably well, and are small enough to easily describe. However, the workload described is fairly particular and mixes several optimizations in single tasks. Thus, we also show MANIMAL performance on a number of synthetic per-optimization tasks.

## C. ADDITIONAL ANALYSIS

In this section we provide additional details on the **analyzer**, in particular how it works with projection and data compression.

Optimizing for **projections** means enumerating which fields of the `map()`'s inputs are never used. We only care

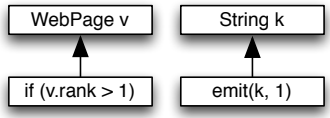

**Figure 5:** Some use-def chains for the `map()` function in Section 2. Nodes represent instructions or variable definitions; each edge's source requires data from its target. We show Java statements for clarity, but the actual graph is computed on bytecodes.

```
1: function findProject(mapperStmts, paramFields):
2:   allPaths ← {}, usedFields ← {}
3:   for s ∈ mapperStmts do
4:     if isEmit(s) then
5:       for all path ∈ paths(s) do
6:         allPaths ← allPaths ∪ conds(path) ∪ s
7:   for all path ∈ allPaths do
8:     for all stmt ∈ path do
9:       usedFields ← usedFields ∪
                      fieldsIn(getUseDef(stmt))
10:  return paramFields − usedFields
```

**Figure 6:** The analyzer detection algorithm for projection. The *paramFields* enumerates all the fields from the serialized `key`, `value` pair. The function *fieldsIn(useDefChain)* returns the parameter fields that appear in the passed-in series of statements.

about calls to `emit()` and control-flow decisions that lead up to `emit()` calls. Other reasons to use inputs – log messages, debugging text, *etc*– we optimize away. The algorithm for projection is seen in Figure 6.

For the sake of space, we discuss **compression**-related techniques only briefly. For direct-operation, **analyzer** first obtains a list of input parameters that are actually used in *map()*. Input parameters for which all uses are equality tests are suitable for direct-operation on compressed data. Finding opportunities for delta-compression is straightforward: **analyzer** simply tests whether the serialized `key` and `value` inputs to `map()` contain numeric values. If so, delta-compression can be applied to those fields.

## D. ADDITIONAL EXPERIMENTS

Here we discuss additional single-optimization results, first presented in Section 4.3. Table 3 shows detailed selection results. We also discuss projection and compression. We tested them using data defined in Figure 7, which is modeled after that used in Pavlo, *et al.*. For these single-optimization experiments, we changed the test data slightly to allow for more interesting queries and a simpler presentation.

For `WebPages` data, we randomly generated unique pages with Zipfian popularity and created the link structure accordingly. The total size of the test `WebPages` file is 129.5 GB. The `UserVisits` data has fields that are all uniformly picked at random from real-world data sets, with the exception of `destURL`. That field was picked from the `WebPages` list of randomly generated URLs (again, according to a Zipfian distribution). The format of this data is nearly the same as with Pavlo, *et al.*, with a few minor typing differences for ease of implementation. We tested the system with 123.7GB of `UserVisits` data.

```
WebPages (
  String url;
  int rank;
  String content);

UserVisits (
  String sourceIP;
  String destURL;
  long visitDate;
  int adRevenue;
  String userAgent;
  String countryCode;
  String languageCode;
  String searchWord;
  int duration;);
```

**Figure 7:** SQL-formatted description of the two kinds of test data we generate to evaluate MANIMAL.

Finally, we represented all of the raw input data in a binary, not textual, format. All runs, whether standard Hadoop or MANIMAL, use this binary format.

### D.0.1 Projection

The projection query consists of the following:

```
SELECT destURL, pageRank from
FROM WebPages WHERE pageRank > threshold
```

Our projection results in Table 4 show that simply removing unneeded serialized fields can have a huge impact on MapReduce job performance. We ran it for three configurations: a Large configuration in which the average WebPage `content` field is 10K (and the total file size is over 123GB), and two Small configurations in which the average `content` field is just 510 bytes (and the total file size is just under 20GB). Of course, for real-life Web content, the Large configuration is much closer to reality.

We ran two different versions of Small in order to avoid distorting effects of Hadoop startup time. Small-1 has the same number of tuples as Large, but its runtime is so small that typical Hadoop startup periods (which can be up to 15 seconds) may have an outsized impact on the result. Small-2 increases the number of documents, and thus the runtime. The ideal solution be to run a Large experiment with the same number of documents as Small-2, but storage constraints prevented us from doing this.

The query does not use `content` in either case, and so it is projected away; the point in using multiple sizes is to show the relative impact of projection optimization when the percentage of optimized-away data is changed. We can see here that even in the Small-1 case, we obtain a 2.4x speedup (and Small-2 is somewhat higher). In the much more realistic Large case, the speedup is more than 27x.

### D.0.2 Compression

As discussed in Section 3, MANIMAL pursues two strategies for improved performance via data compression. First, MANIMAL performs delta-compression on relevant numeric fields in the data. Second, it operates directly on compressed data when program semantics allow it. We can now show runtime results for each of these cases. Note that both of these approaches are orthogonal to builtin MapReduce or filesystem compression mechanisms.

| Selectivity | 60% | 50% | 40% | 30% | 20% | 10% |
|---|---|---|---|---|---|---|
| Intermediate output size | 8.6GB | 7.2GB | 5.8GB | 4.3GB | 2.9GB | 1.4GB |
| Final output size | 72KB | 60KB | 48KB | 36KB | 24KB | 12KB |
| Hadoop Running Time (secs) | 2,004.9 | 1,971.12 | 1,982.80 | 1,995.16 | 1,977.27 | 1,966.94 |
| Manimal **Running Time (secs)** | **1,265.13** | **1,064.69** | **867.91** | **669.09** | **471.66** | **276.72** |
| **Speedup** | **1.59** | **1.85** | **2.29** | **2.98** | **4.19** | **7.10** |

Table 3: Selection times for Manimal on the WebPage data at various levels of selectivity. The indexed input size is 129.5GB.

|  | Small-1 | Small-2 | Large |
|---|---|---|---|
| Original file size | 8.13GB | 19.72 GB | 123.63 GB |
| Number tuples | 11.1M | 27M | 11.1M |
| Average content field size | 510 bytes | 510 bytes | 10K |
| Index size | 743.2 MB | 1.76 GB | 743.2 MB |
| Hadoop running time (secs) | 78.1 | 216.8 | 1,473.8 |
| Manimal running time (secs) | **32.5** | **72.2** | **52.9** |
| Speedup | **2.4** | **3** | **27.8** |

Table 4: Projecting out irrelevant columns allows Manimal to complete a job while processing very few bytes. Creating the index for this task took just slightly longer than running the original task.

One compression strategy we did *not* pursue is traditional dictionary-style compression of the overall input file. Hadoop MapReduce already offers automatic compression and decompression of its inputs, which does not require any semantic insight provided by the Manimal analyzer. All of our experiments show performance increases beyond those enabled by Hadoop's built-in compression support.

We ran a MapReduce program that sums all `duration` values from `UserVisits`. It groups these sums by `destURL`, but does not in the end emit the URL; it simply uses `destURL` as the `key` parameter to `reduce()`. Thus, most of the program does not actually need the true `destURL` value to operate. We could compress `destURL` and leave it compressed, while still retaining the same group-by-`destURL` behavior.

**Delta Compression** We attempted to perform the same experiment using only delta compression on numeric values in the input. Delta compression takes advantage of the fact that sequential data items generally have numeric values that only change slightly. By storing *deltas* instead of the original values, and by employing a size-sensitive encoding that uses fewer bytes to store smaller values, delta compression can dramatically reduce the amount of space needed to store numeric datasets.

We applied delta compression to the `UserVisits` data and ran the query described above. The results appear in Table 5. In order to more clearly show the impact of delta compression, we projected out all non-numeric fields; this post-projection size is listed in the second row of the table. We then delta-compressed `visitDate`, `adRevenue`, `duration`.

Delta compression gives a large space savings (roughly 47%), but yields only a moderate performance boost. However, we note that results described by Abadi, *et al.* [1] on compressing a columnar relational system only show performance improvements when space savings are much larger than those we obtain here. Overall, we regard delta compression as an acceptable, though not spectacular, potential optimization.

|  | Hadoop | Manimal |
|---|---|---|
| Original file size | 123.65 GB | 123.65 GB |
| Post-projection size | 20.99 GB | 20.99 GB |
| Input size (delta-compression) | 20.99 GB | 11.05 GB |
| Running time (secs) | 935.6 | **892.6** |
| Speedup |  | **1.05** |

Table 5: Delta compression on numeric data yields a 47% space savings over the uncompressed version, yielding a moderate performance boost. While delta compression does reduce the amount of bytes that need to be consumed by map(), that function's computational effort is if anything slightly increased, and the shuffle and reduce() loads remain unchanged.

**Operating on Compressed Data** We configured an experiment to use dictionary compression on *only* the `destURL` field prior to processing; all other fields remained uncompressed. During actual program execution, `destURL` is implemented as an integer instead of a `String`. As can be seen in Table 6, Manimal obtains a roughly 2.3x speedup over conventional Hadoop MapReduce. These speedups come from several sources: reduced input size, reduced intermediate data, and faster sorting. We were unable to find opportunities for this optimization in our test set, but the speedup that it permits is substantial; eventually, we would like to examine a larger set of MapReduce programs to see if this optimization can be broadly applied.

## E. FUTURE WORK

An obvious avenue of future work for Manimal is to examine additional optimization techniques, many of which have been mentioned in this paper. In particular, we have done preliminary work on extending analysis beyond the

395

|  | Hadoop | MANIMAL |
| --- | --- | --- |
| Original file size | 123.65 GB | 123.65 GB |
| Indexed file size | 123.65 GB | 76.87 GB |
| Running time (secs) | 4,048 | **1,727** |
| Speedup |  | **2.34** |

Table 6: **Operating on compressed data allows MANIMAL to execute aggregation-style MapReduce programs with an approximately 2.3x speedup over standard Hadoop MapReduce.**

*map* phase. We note that the combined map-shuffle-reduce sequence is akin to a *GROUPBY* query, with the map's output key as the *GROUPBY* value. When results from the *reduce* function are filtered with a conditional clause, the user's program resembles a *GROUPBY* with a *WHERE* clause. If we could accurately predict which temporary *map* outputs will be removed by the *WHERE-related* filtering clause inside *reduce*, then we could delete this temporary data prior to shuffle-reduce without any impact on final program output. We have implemented some infrastructure to perform these optimizations, but performance results are still inconclusive.

Another promising area is to extend MANIMAL techniques to optimize *processing pipelines*. One common form of pipeline is chained MapReduce jobs, in which the output of a given job forms the input of a separate job. One potential difficulty is in simply detecting that two jobs are chained together. However, assuming we can detect the link, it should be quite possible to track relational-style operations across jobs.

Even more ambitious would be to detect *heterogeneous pipelines* that are not strictly limited to MapReduce programs. For example, there might be a Web crawler written in C that feeds content to a MapReduce job, which then yields output that is analyzed with a Python program. We believe these heterogeneous pipelines may be quite common in practice; for example, consider all the data streams and processing that contribute to Web ad pricing or financial trading decisions. For large enough processing tasks, it is probably unreasonable to ever hope that the entire processing stream will ever be rewritten using a single language or framework. We believe they make a very exciting topic for future investigation.